\documentclass[%
 reprint,
 amsmath,amssymb,
 aps,
 prb,
 floatfix,
]{revtex4-2}

\usepackage{graphicx}
\usepackage[hidelinks]{hyperref}
\usepackage{siunitx}
\usepackage{tikz}
\usetikzlibrary{shapes.geometric, arrows.meta, positioning}

\begin{document}

\title{A transferable full-band Monte Carlo framework for complex alloy avalanche photodiodes}

\author{Shafat Shahnewaz}
\thanks{These authors contributed equally to this work.}
\affiliation{Department of Electrical and Computer Engineering, University of Virginia, Charlottesville, Virginia 22903, USA}
\author{Hannaneh Karimi}
\thanks{These authors contributed equally to this work.}
\affiliation{Department of Electrical and Computer Engineering, University of Virginia, Charlottesville, Virginia 22903, USA}
\author{Joe C. Campbell}
\affiliation{Department of Electrical and Computer Engineering, University of Virginia, Charlottesville, Virginia 22903, USA}
\author{Avik W. Ghosh}
\email{ag7rq@virginia.edu}
\affiliation{Department of Electrical and Computer Engineering, University of Virginia, Charlottesville, Virginia 22903, USA}
\affiliation{Department of Physics, University of Virginia, Charlottesville, Virginia 22903, USA}

\date{September 30, 2026}

\begin{abstract}
We present a physics-based multiscale full-band Monte Carlo framework for modeling avalanche multiplication and excess noise in complex alloy avalanche photodiodes (APDs), demonstrated on Al$_{0.7}$InAsSb as a representative quaternary system. The framework links atomistic material structure to device-level avalanche statistics: an environment-dependent $sp^3d^5s^\ast$ tight-binding calculation resolves the full conduction- and valence-band structure of the random- or digital-alloy configuration -- including $\Gamma$, X, and L valley ordering, non-parabolicity, anisotropy, and spin-orbit-induced valence-band splitting -- and supplies the band-structure inputs for stochastic high-field transport. A central element of the framework is a physics-derived treatment of alloy-disorder scattering, in which the quaternary disorder potential is constructed from atomic valence differences, covalent radii, and Thomas--Fermi screening through a composition-weighted decomposition into binary contributions, complemented by composition-interpolated polar-optical, acoustic, intervalley-phonon, impurity, and impact-ionization models. Because every material-dependent input is generated from the atomic composition and configuration by the same well-defined procedure, the framework transfers without structural modification to arbitrary zinc-blende ternary and quaternary alloys. Applied to a \SI{1}{\micro\meter} Al$_{0.7}$InAsSb p-i-n APD, the framework reproduces the measured gain and excess-noise characteristics with only the impact-ionization softness parameters calibrated. The approach provides a documented, reproducible route for analyzing and designing complex alloy APDs in which band structure, disorder, and scattering physics jointly determine gain and ionization statistics.
\end{abstract}

\maketitle

\section{Introduction}
Avalanche photodiodes (APDs) use impact ionization to provide internal gain for weak optical signals. The same stochastic multiplication process also introduces excess noise. In the local-field model of McIntyre, the excess-noise factor is controlled mainly by the ratio of hole and electron ionization coefficients~\cite{mcintyre1966multiplication}. Nonlocal dead-space models become important when the multiplication region is thin enough that the carrier energy history affects the ionization statistics~\cite{hayat1992deadspace,saleh2001impact}. Sb-based alloys provide an additional motivation for a microscopic transport description. AlInAsSb APDs have shown low ionization-coefficient ratios and low excess noise~\cite{woodson2016lownoise,jones2020lownoise,dadey2023considerations}. Deviations from a simple local-field description have also been reported in relatively thick Sb-based multiplication regions~\cite{lewis2023anomalous,karimi2025analysis}. These results make AlInAsSb a useful test system for models that resolve both the electronic structure and the competing scattering processes.

\begin{figure*}[t]
\includegraphics[width=0.9\linewidth]{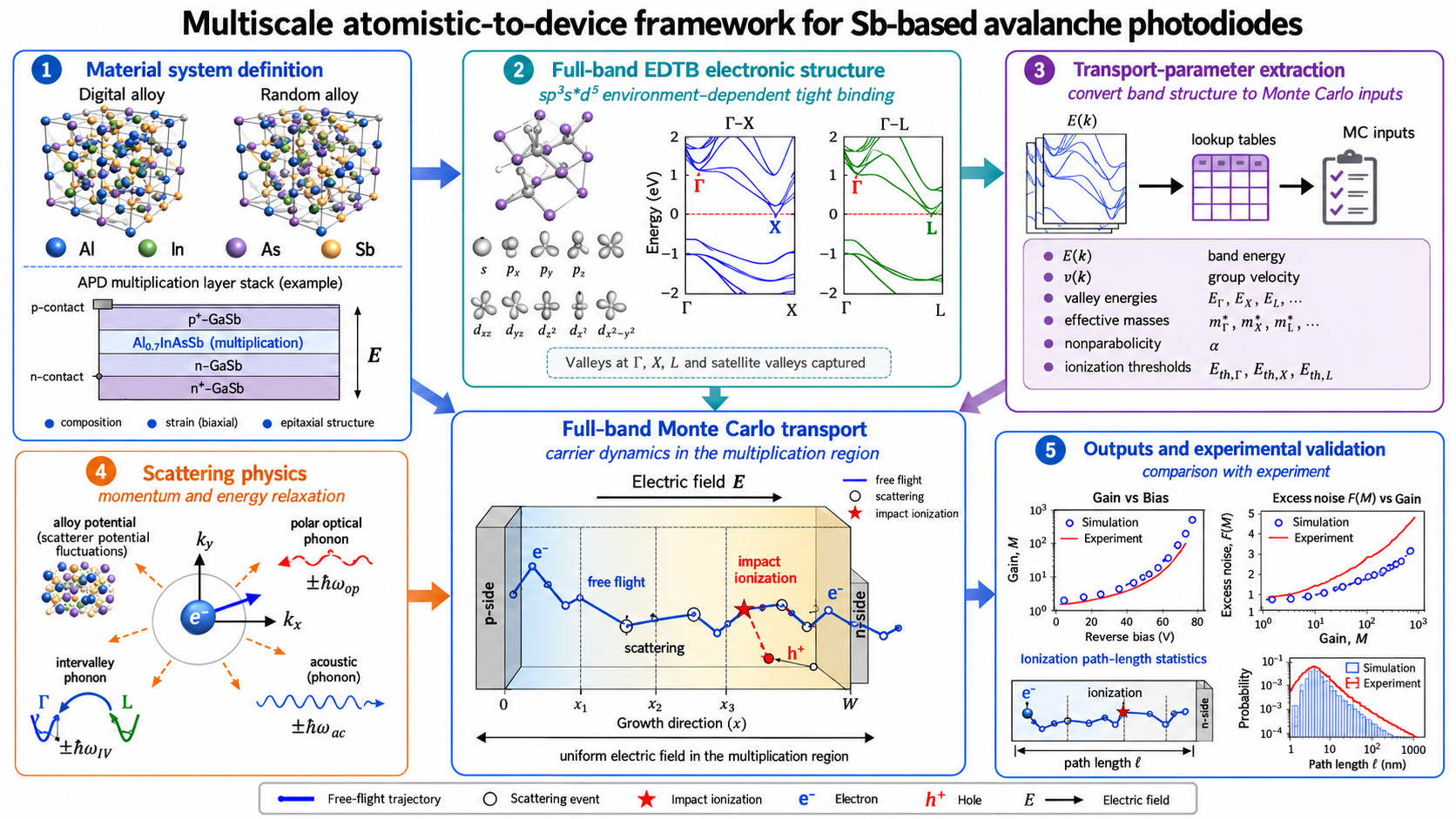}
\caption{High-level workflow of the multiscale framework. The device material structure is first defined from the APD layer design and atomistic alloy configuration. Environment-dependent tight-binding calculations are then used to extract band-structure-derived Monte Carlo inputs, which are combined with composition-derived scattering models and calibrated impact-ionization physics to simulate avalanche transport and extract gain and excess-noise characteristics.The schematic was generated with assistance from OpenAI ChatGPT (GPT-5.6) based on author-provided scientific content and was reviewed and verified by the authors.}
\label{fig:flowchart}
\end{figure*}

The modeling problem is challenging because the relevant material inputs change with alloy composition and growth sequence. High-Al-content Al$_x$In$_{1-x}$As$_y$Sb$_{1-y}$ can be grown as a digital alloy using short-period stacks of binary layers, including compositions within the random-alloy miscibility gap~\cite{maddox2016broadly}. The resulting electronic structure contains folded bands and mini-gaps that are not captured by a small set of bulk effective masses and valley offsets. Full-band Monte Carlo transport can retain these features, but it requires a band structure and a consistent set of scattering rates over the carrier-energy range relevant to avalanche transport~\cite{jacoboni1983monte,shichijo1981band}.

Earlier studies provide important parts of this framework. Littlejohn \textit{et al.} developed alloy-scattering expressions for ternary and quaternary III--V materials~\cite{littlejohn1978alloy}. Tan \textit{et al.} introduced a transferable environment-dependent $sp^3d^5s^\ast$ tight-binding model for strained group-IV and III--V materials and heterostructures~\cite{tan2016transferable}. Zheng \textit{et al.} combined this electronic structure with full-band Monte Carlo transport for AlInAsSb digital alloys and showed that the conduction-band mini-gap does not prevent electron impact ionization~\cite{zheng2020fullband}. More recent AlInAsSb Monte Carlo work showed that the relative strengths of alloy and phonon scattering have a strong effect on the calculated multiplication noise~\cite{karimi2025analysis}. A screened atomic-potential approach for estimating alloy scattering in Sb-based quaternaries was also reported in our earlier work~\cite{shahnewaz2025alloy}.

Here we combine these ingredients into one modeling workflow. The atomistic alloy structure is first mapped to an environment-dependent tight-binding Hamiltonian. The calculated bands provide valley energies, carrier velocities, effective masses, and nonparabolicity used by the transport model. Composition-dependent scattering models are then combined with a Keldysh impact-ionization rate in a Monte Carlo solver. We demonstrate the procedure for Al$_{0.7}$InAsSb and compare the calculated gain and excess noise with measurements from a \SI{1}{\micro\meter} p-i-n APD. The focus of this paper is the construction and validation of the framework rather than a unique microscopic assignment of the low-noise mechanism.

\section{Modeling framework}
The framework proceeds in four stages, summarized in Fig.~\ref{fig:flowchart}. First, the atomistic material structure is defined for the APD multiplication region. Second, the electronic structure is calculated with environment-dependent tight binding (EDTB). Third, alloy-disorder and phonon-scattering models are parameterized for the chosen composition. Finally, the resulting band and scattering tables are used in a Monte Carlo avalanche-transport calculation with a calibrated impact-ionization model. The same sequence can be reused for another alloy once its atomistic structure and required constituent parameters are supplied.

\subsection{Environment-dependent tight-binding electronic structure}

The multiplication-region material is modeled as Al$_x$In$_{1-x}$As$_y$Sb$_{1-y}$, with Al$_{0.7}$InAsSb used as the representative quaternary alloy. High-Al-content AlInAsSb has been grown as a digital alloy to access compositions inside the random-alloy miscibility gap~\cite{maddox2016broadly}. In the structure considered here, one digital-alloy period contains 3 monolayers (ML) of InAs, 3 ML of AlSb, 1 ML of AlAs, and 3 ML of AlSb on a GaSb substrate, following Ref.~\cite{zheng2020fullband}. The atomistic description retains the local bonding, interfaces, and strain that modify the valley alignment and band curvature relevant to high-field transport.

\begin{figure*}[t]
\includegraphics[width=0.7\linewidth]{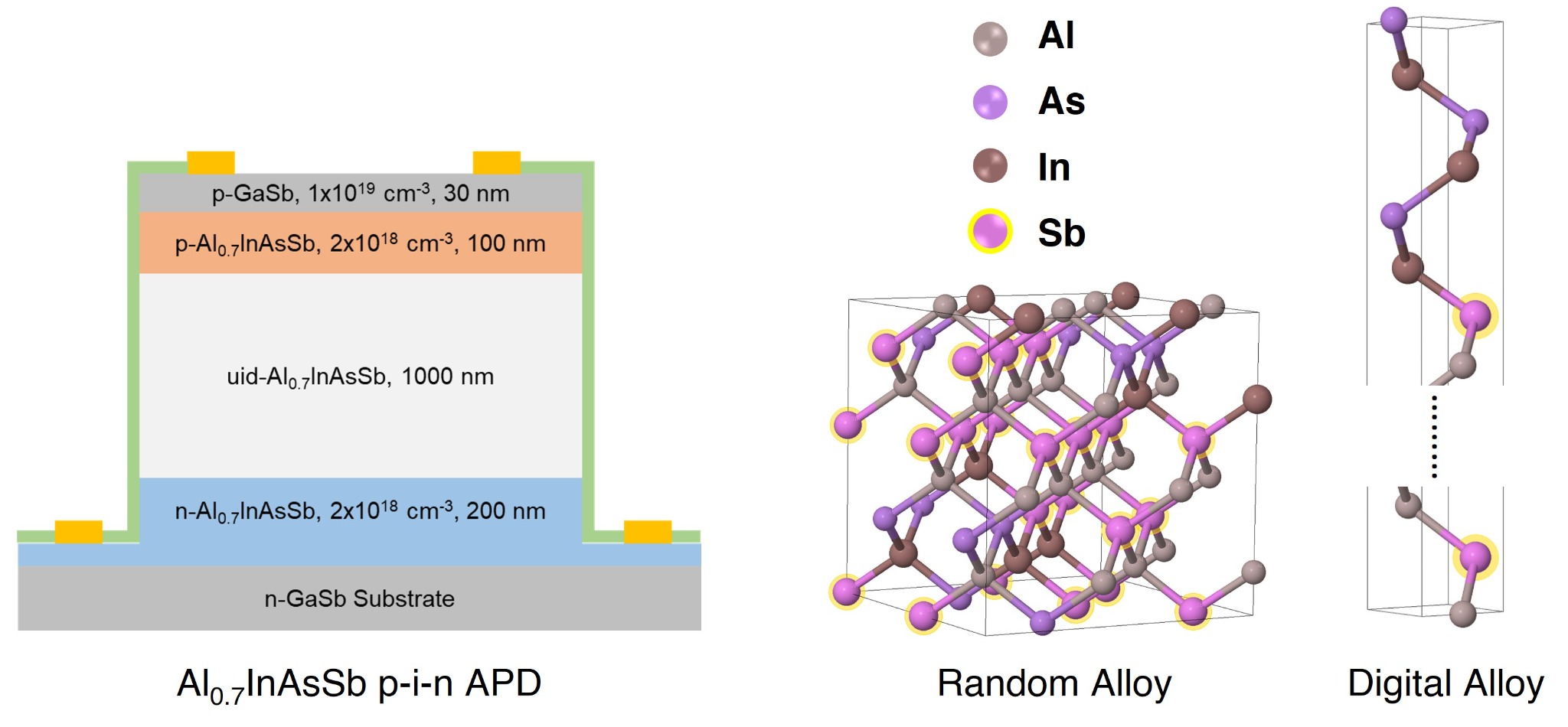}
\caption{Schematic cross-section of the Al$_{0.7}$InAsSb p-i-n APD structure used throughout this work, where the unintentionally doped (UID) region can be implemented as either a digital or a random alloy.}
\label{fig:pin}
\end{figure*}

The electronic structure is computed using an environment-dependent $sp^3d^5s^\ast$ tight-binding (EDTB) model~\cite{tan2016transferable,zheng2020fullband}. Each atom is represented by an orbital basis containing one $s$, three $p$, five $d$, and one excited $s^\ast$ orbital per spin, allowing the model to describe both the conduction-band valley structure and the valence-band splitting relevant to impact ionization~\cite{tan2016transferable}. The tight-binding basis state is denoted by $\left|\psi_{\alpha i}\right\rangle$, where $i$ labels the atom and $\alpha$ labels the orbital. The single-particle wave function is expanded as
\begin{equation}
\left|\Psi_{n\mathbf{k}}\right\rangle =
\sum_{i,\alpha} C^{(n)}_{i\alpha}(\mathbf{k})
e^{i\mathbf{k}\cdot \mathbf{R}_i}
\left|\psi_{\alpha i}\right\rangle ,
\label{eq:tb_wavefunction}
\end{equation}
where $n$ is the band index, $\mathbf{k}$ is the crystal wave vector, and $\mathbf{R}_i$ is the atomic position. The band energies are obtained from the eigenvalue problem
\begin{equation}
\sum_{j,\beta}
H_{\alpha i,\beta j}(\mathbf{k})
C^{(n)}_{j\beta}(\mathbf{k})
=
E_n(\mathbf{k})
C^{(n)}_{i\alpha}(\mathbf{k}).
\label{eq:tb_eigenproblem}
\end{equation}

The key feature of the EDTB model is that the Hamiltonian parameters are determined by the local atomic environment rather than by a single bulk material label~\cite{tan2016transferable}. This is particularly important for AlInAsSb digital alloys and heterointerfaces, where the same atomic species may have different nearest-neighbor configurations depending on whether it is located in an AlSb-like, InAs-like, AlAs-like, or interfacial bonding environment. Following the environment-dependent formulation, the local potential near atom $i$ is written as the sum of the central atomic potential and the potentials from neighboring atoms,
\begin{equation}
U_i^{\mathrm{tot}}(\mathbf{r})
=
U_i(|\mathbf{r}|)
+
\sum_{j\in \mathrm{NN}}
U_j(|\mathbf{r}-\mathbf{d}_{ij}|),
\label{eq:local_potential}
\end{equation}
where $\mathbf{d}_{ij}$ is the vector from atom $i$ to a nearest neighbor $j$. The neighbor contribution is expanded in multipoles,
\begin{equation}
U_j(|\mathbf{r}-\mathbf{d}_{ij}|)
=
\sum_l U^{(l)}_j(r,d_{ij})
\sum_{m=-l}^{l}
Y_{lm}^{\ast}(\hat{\mathbf{r}})
Y_{lm}(\hat{\mathbf{d}}_{ij}),
\label{eq:multipole_expansion}
\end{equation}
so that hydrostatic bond-length changes and lower-symmetry strain or interface perturbations enter the tight-binding Hamiltonian in a systematic way. Retaining multipoles up to quadrupole order gives
\begin{equation}
H = H^{(0)} + H^{(1)} + H^{(2)},
\label{eq:edtb_hamiltonian}
\end{equation}
where $H^{(0)}$ contains the ordinary two-center tight-binding terms and hydrostatic environment corrections, while $H^{(1)}$ and $H^{(2)}$ describe dipole- and quadrupole-induced corrections arising from bond-angle changes, strain, and local symmetry breaking.

The diagonal on-site matrix elements contain both an atom-specific orbital energy and environment-dependent shifts from neighboring atoms:
\begin{equation}
H^{(0)}_{\alpha i,\alpha i}
=
E_{\alpha i}
+
\sum_{j\in \mathrm{NN}} I_{\alpha i,j}(d_{ij})
+
\sum_{j\in \mathrm{NN}} O_{i,j}(d_{ij}),
\label{eq:onsite}
\end{equation}
with
\begin{equation}
I_{\alpha i,j}(d_{ij})
=
I_{\alpha i,j}
\exp[-\lambda_{\alpha i,j}(d_{ij}+\delta d_{ij}-d_0)],
\label{eq:I_environment}
\end{equation}
and
\begin{equation}
O_{i,j}(d_{ij})
=
O_{i,j}
\exp[-\lambda_{i,j}(d_{ij}+\delta d_{ij}-d_0)].
\label{eq:O_environment}
\end{equation}
Here $d_0$ is a reference bond length, $I_{\alpha i,j}$ describes orbital-dependent environment corrections, and $O_{i,j}$ accounts for orbital-independent band-offset contributions. The parameter $\delta d_{ij}$ is used in the transferable parametrization to align the tight-binding band edges with the target room-temperature band structure. Spin-orbit coupling is also included in the on-site block, which is essential for Sb-containing compounds because the large spin-orbit interaction strongly affects the split-off band and the valence-band density of states.

The off-site coupling between orbital $\alpha$ on atom $i$ and orbital $\beta$ on atom $j$ follows Slater-Koster symmetry with explicit bond-length dependence,
\begin{equation}
V_{\alpha i,\beta j}^{|m|}(d_{ij})
=
V_{\alpha i,\beta j}^{|m|}
\exp[-\eta_{\alpha i,\beta j}^{|m|}(d_{ij}+\delta d_{ij}-d_0)],
\label{eq:bond_scaling}
\end{equation}
where $|m|=\sigma,\pi,\delta$ specifies the angular character of the two-center interaction. The angular dependence of the hopping matrix is then constructed from the bond direction cosines through the Slater-Koster tables. In addition, the dipole and quadrupole terms in Eq.~\eqref{eq:edtb_hamiltonian} generate off-diagonal on-site and interatomic corrections when the local symmetry is broken by strain or heterointerfaces. Thus, the same Hamiltonian can be applied to random alloys, digital alloys, and short-period superlattices without redefining the material identity at each interface atom.

For a periodic digital alloy, the Bloch Hamiltonian is assembled as
\begin{equation}
H_{\alpha i,\beta j}(\mathbf{k})
=
\sum_{\mathbf{R}}
H_{\alpha i,\beta j}(\mathbf{R})
e^{i\mathbf{k}\cdot(\mathbf{R}+\mathbf{r}_j-\mathbf{r}_i)},
\label{eq:bloch_hamiltonian}
\end{equation}
where $\mathbf{R}$ runs over neighboring unit cells and $\mathbf{r}_i$ and $\mathbf{r}_j$ are basis-atom positions inside the supercell. Diagonalization of Eq.~\eqref{eq:tb_eigenproblem} over the Brillouin zone yields the full-band dispersion $E_n(\mathbf{k})$ and eigenvectors. The computed band structure resolves the $\Gamma$, X, and L conduction valleys, the heavy-hole, light-hole, and split-off valence bands, and the mini-gaps introduced by the short-period digital-alloy potential.

\begin{figure*}[t]
\includegraphics[width=0.8\linewidth]{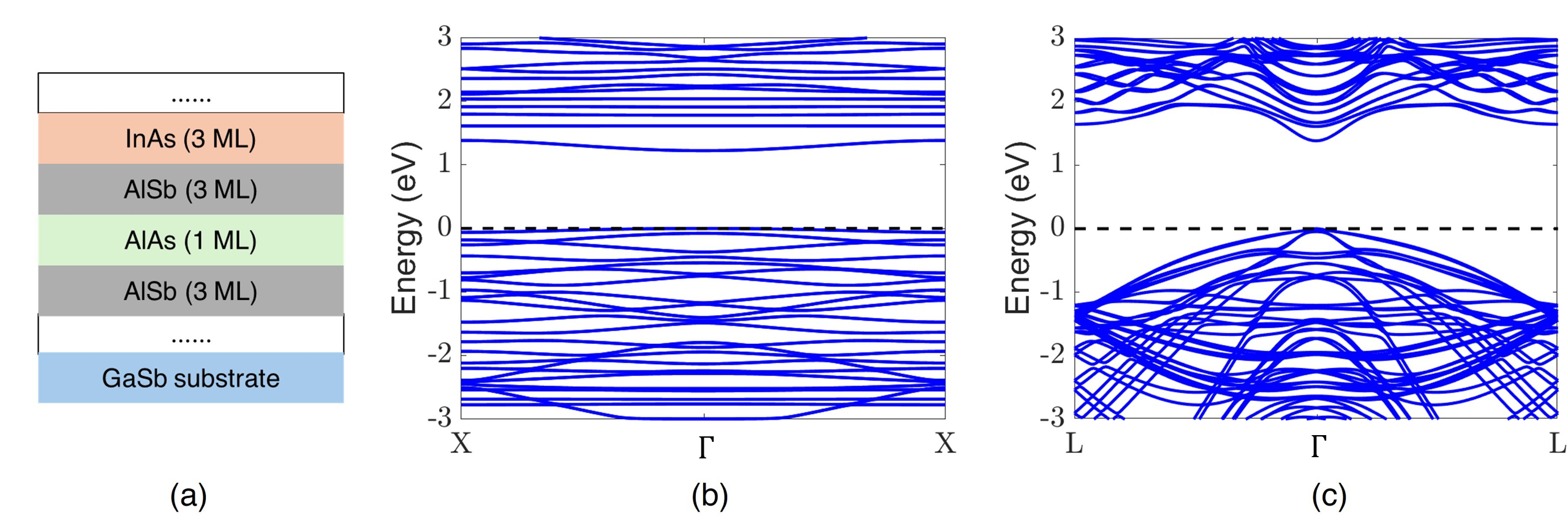}
\caption{(a) Digital alloy superlattice structure of Al$_{0.7}$InAsSb grown on a GaSb substrate. (b) Band structure along the $\Gamma$--X direction and (c) along the $\Gamma$--L direction, showing the electronic states used to construct the full-band Monte Carlo transport inputs.}
\label{fig:bandstructure}
\end{figure*}

The EDTB calculation provides the band-structure quantities required by the Monte Carlo solver. The carrier group velocity is obtained directly from the full-band dispersion,
\begin{equation}
\mathbf{v}_n(\mathbf{k})
=
\frac{1}{\hbar}\nabla_{\mathbf{k}}E_n(\mathbf{k}),
\label{eq:group_velocity}
\end{equation}
and the effective-mass tensor near a valley extremum is computed from the band curvature,
\begin{equation}
\left(m_n^{-1}\right)_{ij}
=
\frac{1}{\hbar^2}
\frac{\partial^2 E_n(\mathbf{k})}{\partial k_i \partial k_j}.
\label{eq:mass_tensor}
\end{equation}
The valley offsets are extracted by locating the minima of the conduction-band valleys in $k$-space:
\begin{equation}
\Delta E_{\Gamma X}=E_X-E_\Gamma,
\qquad
\Delta E_{\Gamma L}=E_L-E_\Gamma .
\label{eq:valley_offsets}
\end{equation}
These offsets determine the energetic accessibility of intervalley transfer during high-field transport. Nonparabolicity is represented either by direct interpolation of the full $E_n(\mathbf{k})$ tables or, for compact scattering-rate parameterization, by a Kane-like relation,
\begin{equation}
E(1+\alpha E)=\frac{\hbar^2 k^2}{2m^\ast},
\label{eq:nonparabolicity}
\end{equation}
where $\alpha$ is the nonparabolicity parameter. In the present work, carrier drift, valley assignment, and impact-ionization eligibility are evaluated from the full-band tables, while the analytic scattering-rate models of Sec.~\ref{sec:scattering} are parameterized by the EDTB-derived effective masses and nonparabolicity parameters.

\begin{table}[t]
\centering
\caption{EDTB-derived band-structure parameters of Al$_{0.7}$InAsSb used in the Monte Carlo model.}
\label{tab:bandparams}
{
\renewcommand{\arraystretch}{1.2}
\begin{ruledtabular}
\begin{tabular}{lcc}
Valley & Band edge (eV) & Effective mass ($m^\ast/m_0$) \\
\hline
$\Gamma$ & 1.19 & 0.07 \\
$L$ & 1.71 & 0.10 \\
$X$ & 1.51 & 1.30 \\
Heavy hole & 0 & 0.35 \\
Light hole & 0 & 0.07 \\
Split-off & 0.65 & 0.17 \\
\end{tabular}
\end{ruledtabular}
}
\end{table}

The values in Table~\ref{tab:bandparams} are used as compact valley-resolved inputs to the Monte Carlo transport model. The $\Gamma$ valley defines the electron injection edge and the band gap, while the X and L valleys provide high-density final states for intervalley transfer. The X valley lies close enough to the $\Gamma$ valley to become accessible during high-field acceleration, and its large effective mass reflects the high density of states available for scattering. The valence-band parameters determine the hole acceleration pathways and the ionization threshold for hole-initiated multiplication. The split-off band is especially important in Sb-based alloys because spin-orbit coupling separates it substantially from the heavy- and light-hole bands, thereby modifying the valence-band density of states and the phase space for hole impact ionization.

A full-band description is needed because the digital-alloy superlattice produces mini-gaps and avoided crossings that are not represented by a single parabolic band. Prior AlInAsSb calculations showed that electrons can reach in-plane states that bypass the apparent conduction-band mini-gap along the growth direction~\cite{zheng2020fullband}. The valence bands have different masses and a sizable split-off separation, which changes the high-energy phase space available to holes. These features motivate retaining the three-dimensional band structure in the transport model.

To interface the EDTB electronic structure with the Monte Carlo solver, the first Brillouin zone is discretized on a three-dimensional $k$ mesh. The band energy and group velocity are stored as lookup tables,
\begin{equation}
\{E_n(\mathbf{k}),\mathbf{v}_n(\mathbf{k}),\nu_n(\mathbf{k})\},
\label{eq:mc_lookup}
\end{equation}
where $\nu_n(\mathbf{k})$ denotes the valley assignment used by the transport model. These tables retain the full-band dispersion for carrier drift and valley tracking. Compact quantities such as valley offsets, effective masses, and nonparabolicity parameters are extracted from the same EDTB calculation for the analytic scattering-rate expressions below.

\section{Composition-derived scattering models}
\label{sec:scattering}
The scattering rates entering the transport solver are based on Fermi's golden rule and standard semiconductor Monte Carlo formulations~\cite{lundstrom2000fundamentals,jacoboni1983monte,jacoboni1989monte,tomizawa1993numerical}. The EDTB calculation supplies band quantities such as valley separations, effective masses, and nonparabolicity. Alloy-disorder parameters are evaluated from the alloy composition and the screened atomic-potential model described below. Phonon parameters are taken from, or interpolated from, the binary constituents as specified for each mechanism.

\subsection{Alloy-disorder scattering}
\label{sec:alloy}
In a substitutional III--V alloy, fluctuations in the local crystal potential produce alloy-disorder scattering~\cite{littlejohn1978alloy}. We use the composition-weighted quaternary form together with the screened atomic-potential estimate introduced in our earlier work~\cite{shahnewaz2025alloy}. The alloy scattering rate is written as
\begin{equation}
\frac{1}{\tau_{\mathrm{alloy}}}=\frac{3\pi}{8\sqrt{2}}\frac{(m^*)^{3/2}}{\hbar^4}\sqrt{\gamma(E)}\frac{d\gamma}{dE}\Omega |\Delta U_Q|^2 S,
\label{eq:alloy}
\end{equation}
where $m^*$ is the effective mass, $\Omega=a^3/4$ is the primitive cell volume, and $S$ is the ordering parameter ($S=1$ for a fully random alloy). The non-parabolic dispersion is incorporated through
\begin{equation}
\gamma(E)=E(1+\sigma E), \qquad \sigma=\frac{1}{E_g}\left(1-\frac{m^*}{m_0}\right)^2 .
\end{equation}

For a general quaternary alloy $A_xB_{1-x}C_yD_{1-y}$, the total squared disorder potential is obtained through a composition-weighted summation of binary contributions:
\begin{align}
|\Delta U_Q(x,y)|^2 ={}& x(1-x)y^2|\Delta U_{ABD}|^2 \nonumber\\
&+ x(1-x)(1-y)^2|\Delta U_{ABC}|^2 \nonumber\\
&+ x^2y(1-y)|\Delta U_{BCD}|^2 \nonumber\\
&+ (1-x)^2y(1-y)|\Delta U_{ACD}|^2 .
\label{eq:quaternary_deltaU}
\end{align}
Rather than treating the binary disorder potentials as fitting parameters, we derive them from atomic properties. For a ternary alloy $A_xB_{1-x}C$, the disorder potential between species $A$ and $B$ is
\begin{equation}
\Delta U=\frac{b}{4\pi\epsilon_0}\left(\frac{Z_A}{r_A}-\frac{Z_B}{r_B}\right)\exp(-k_sR),
\label{eq:deltaU}
\end{equation}
where $Z$ and $r$ denote the valence number and covalent radius, respectively, $R$ is the effective bond length, and $k_s$ is the Thomas--Fermi screening wave vector. For zinc-blende materials, $b=1.5$. The screening wave vector is evaluated from
\begin{equation}
k_s=\sqrt{\frac{4k_F}{\pi a_B}}, \qquad k_F=(3\pi^2N_{\mathrm{val}})^{1/3},
\end{equation}
with $N_{\mathrm{val}}=32/a^3$ and lattice constant $a$. For a ternary alloy $A_{1-x}B_xC$, the effective bond length is approximated as
\begin{equation}
R=\frac{1}{2}\left(xr_A+(1-x)r_B+r_C\right).
\end{equation}
For Al$_{0.7}$InAsSb, with $A=\mathrm{Al}$, $B=\mathrm{In}$, $C=\mathrm{As}$, and $D=\mathrm{Sb}$, this procedure gives $\Delta U_{ABC}=\SI{0.64}{eV}$, $\Delta U_{ABD}=\SI{0.52}{eV}$, $\Delta U_{ACD}=\SI{2.65}{eV}$, and $\Delta U_{BCD}=\SI{3.01}{eV}$. The larger terms contain the As--Sb contrast and are consistent with reports that Sb-based quaternaries can have strong alloy scattering because of the large covalent-radius mismatch~\cite{guo2022temperature,karimi2025analysis}.

In the Monte Carlo implementation, alloy scattering is treated as elastic. The event changes the carrier momentum but does not introduce an energy-loss term. The rate in Eq.~\eqref{eq:alloy} therefore competes with inelastic phonon processes without directly cooling the carrier distribution.


\subsection{Polar optical phonon scattering}
Polar optical phonon scattering is described by the Fr{\"o}hlich interaction~\cite{frohlich1954electrons,jacoboni1983monte}. The process is inelastic and exchanges the LO-phonon energy $\hbar\omega_{\mathrm{LO}}$ with the carrier. The long-range interaction favors small momentum transfer. An optical phonon energy of \SI{24}{meV} is used in the present Al$_{0.7}$InAsSb parameter set.

\subsection{Intervalley phonon scattering}
Electrons can transfer between the $\Gamma$, X, and L valleys through short-wavelength phonons that supply the required momentum. The process is treated as isotropic and inelastic in the rate model~\cite{jacoboni1983monte,jacoboni1989monte}. For ellipsoidal, nonparabolic valleys, the rate is
\begin{align}
\frac{1}{\tau_{\mathrm{iv}}}={}&\frac{(D_tK)_i^2 m^{3/2}Z_f}{2^{1/2}\pi\rho\hbar^3\omega_i}
\begin{bmatrix}N_i\\N_i+1\end{bmatrix}
\left(\varepsilon\pm\hbar\omega_i-\Delta\varepsilon_{fi}\right)^{1/2} \nonumber\\
&\times \left[1+2\alpha\left(\varepsilon\pm\hbar\omega_i-\Delta\varepsilon_{fi}\right)\right],
\label{eq:iv}
\end{align}
where $D_tK$ is the deformation potential, $m$ is the effective mass, $Z_f$ is the number of final valleys, $\rho$ is the mass density, $\omega_i$ is the intervalley phonon angular frequency, $\alpha$ is the non-parabolicity coefficient, $\Delta\varepsilon_{fi}$ is the energy difference between the final and initial valley minima, and $N_i$ is the Bose--Einstein occupation (absorption: $N_i$; emission: $N_i+1$).

The deformation potential for Al$_{0.7}$InAsSb is obtained by effective-medium Vegard-type interpolation~\cite{obukhov2009ab} over the binary constituents,
\begin{align}
D_{\mathrm{eff}}={}&w_{\mathrm{AlAs}}D_{\mathrm{AlAs}}+w_{\mathrm{AlSb}}D_{\mathrm{AlSb}} \nonumber\\
&+w_{\mathrm{InAs}}D_{\mathrm{InAs}}+w_{\mathrm{InSb}}D_{\mathrm{InSb}}.
\end{align}
with composition weights $w_{\mathrm{AlAs}}=xy$, $w_{\mathrm{AlSb}}=x(1-y)$, $w_{\mathrm{InAs}}=(1-x)y$, and $w_{\mathrm{InSb}}=(1-x)(1-y)$. The valley separations entering $\Delta\varepsilon_{fi}$ are taken directly from the EDTB band structure (Table~\ref{tab:bandparams}); the $\Gamma$--X separation of \SI{0.32}{eV} together with an intervalley phonon energy of \SI{34}{meV} sets the onset of the dominant $\Gamma\!\to\!\mathrm{X}$ channel, whose strength reflects the high X-valley density of states.

\subsection{Acoustic phonon scattering}
Long-wavelength acoustic vibrations are treated with the usual intravalley deformation-potential model. At room temperature we use the equipartition approximation and treat the event as elastic. The acoustic phonon energy is \SI{11}{meV} in the present parameter set~\cite{lundstrom2000fundamentals,jacoboni1983monte}.

\subsection{Ionized impurity scattering}
Ionized impurity scattering is treated as elastic and anisotropic~\cite{conwell1950theory}. Its rate decreases with carrier energy, so it is significant only at low fields and low energies; in the high-field multiplication region it is subdominant but retained for completeness.

\begin{figure}[t]
\includegraphics[width=\columnwidth]{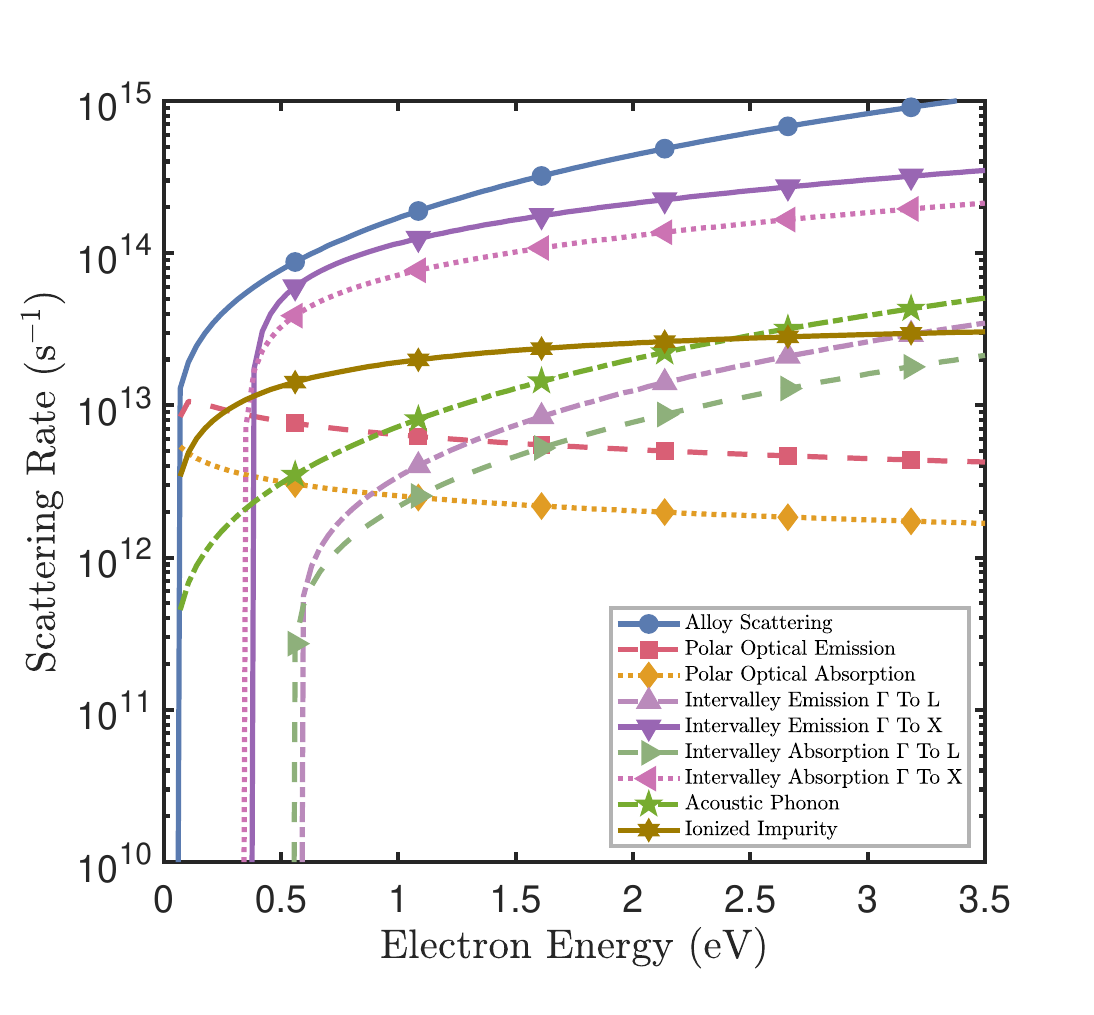}
\caption{Scattering rates in Al$_{0.7}$InAsSb as a function of carrier energy, computed from the composition-derived models of Sec.~\ref{sec:scattering} with EDTB band-structure inputs.}
\label{fig:scattering}
\end{figure}

The resulting electron scattering rates are shown in Fig.~\ref{fig:scattering}. For the present Al$_{0.7}$InAsSb parameter set, alloy scattering is the largest rate over much of the energy range relevant to avalanche buildup. This trend is consistent with earlier Monte Carlo analysis of Sb-based APDs~\cite{karimi2025analysis}.

\section{Monte Carlo transport and calibration}
\label{sec:mc}
\begin{figure}[t]
\centering
\begin{tikzpicture}[
    font=\small,
    node distance=5mm,
    >=Latex,
    startstop/.style={
        ellipse,
        draw=black,
        fill=green!30,
        minimum width=34mm,
        minimum height=10mm,
        align=center
    },
    process/.style={
        rectangle,
        rounded corners=3pt,
        draw=black,
        fill=blue!25,
        minimum width=48mm,
        minimum height=11mm,
        align=center
    },
    decision/.style={
        diamond,
        aspect=2.2,
        draw=black,
        fill=orange!35,
        minimum width=42mm,
        minimum height=16mm,
        align=center,
        inner sep=1pt
    },
    arrow/.style={
        ->,
        thick
    }
]

\node[startstop] (start) {START};

\node[process, below=of start] (tau)
{$\tau =$ free flight time};

\node[process, below=of tau] (drift)
{Drift$(\tau)$};

\node[process, below=of drift] (scatter)
{Scattering process};

\node[decision, below=of scatter] (contact)
{Carrier reaches\\contact?};

\node[startstop, below=of contact] (stop)
{STOP};

\draw[arrow] (start) -- (tau);
\draw[arrow] (tau) -- (drift);
\draw[arrow] (drift) -- (scatter);
\draw[arrow] (scatter) -- (contact);
\draw[arrow] (contact) -- node[right] {Yes} (stop);
\draw[arrow]
(contact.west) -- ++(-1.15,0)
|- node[pos=0.25,left] {No} (tau.west);

\end{tikzpicture}
\caption{Flowchart of the Monte Carlo transport loop. A free-flight time is selected, the carrier is drifted under the applied electric field, and a scattering event is applied. The loop is repeated until the carrier reaches a device contact.}
\label{fig:mc_flowchart}
\end{figure}
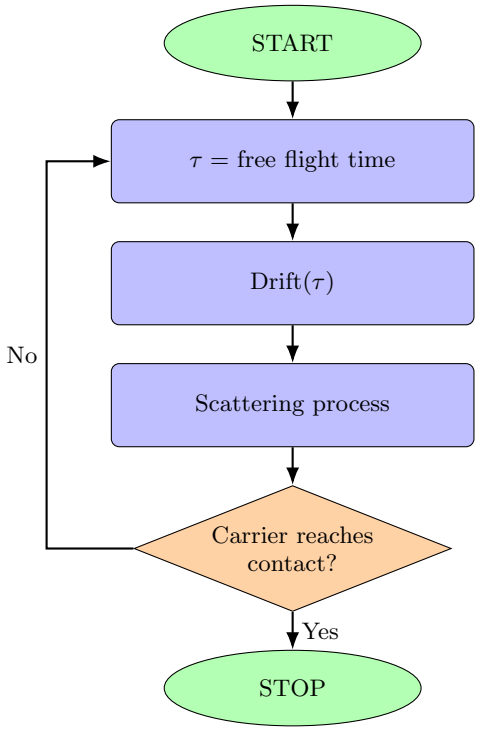

The Monte Carlo simulation follows carriers through a sequence of field-driven free flights and stochastic scattering events~\cite{jacoboni1983monte,jacoboni1989monte}. A carrier is injected into the multiplication region and accelerated by the electric field. The free-flight time is
\begin{equation}
\tau = \frac{-\ln(r)}{\Gamma},
\label{eq:freeflight}
\end{equation}
where $r$ is a uniform random number on $(0,1)$ and $\Gamma$ is the total scattering rate. At the end of each free flight a scattering mechanism is selected with probability proportional to its rate, and the carrier energy and wavevector are updated according to the elastic/inelastic and isotropic/anisotropic character of the selected mechanism described in Sec.~\ref{sec:scattering}. Impact ionization is included as one of the competing scattering channels: a carrier above threshold can excite a valence-band electron into the conduction band, generating a secondary electron-hole pair. The simulation iterates until the primary carrier and all secondaries have exited the UID region (Fig.~\ref{fig:mc_flowchart}).

\subsection{Impact-ionization model and calibration protocol}
The impact-ionization rate is described by the Keldysh formula~\cite{keldysh1960kinetic},
\begin{equation}
\Gamma_{ii}=\Lambda\left(E-E_{th}\right)^{\Upsilon},
\label{eq:keldysh}
\end{equation}
where $E_{th}$ is the threshold energy, $\Lambda$ is the softness parameter, and $\Upsilon$ is the power-law index. The Keldysh parameter set used in this work is listed in Table~\ref{tab:mcparams}. The EDTB band structure and the non-ionizing scattering tables are fixed before the device calculation. The remaining impact-ionization parameters are calibrated to the measured Al$_{0.7}$InAsSb device characteristics. This wording keeps the fitted ionization model separate from the composition-dependent transport inputs.

\begin{table}[t]
\centering
\caption{Keldysh impact-ionization parameters used in the Monte Carlo model.}
\label{tab:mcparams}
\renewcommand{\arraystretch}{1.3}
\begin{ruledtabular}
\begin{tabular}{lc}
Simulation parameter & Value \\
\hline
Electron threshold, $E_{\mathrm{th},e}$ (eV) & 2.1 \\
Hole threshold, $E_{\mathrm{th},h}$ (eV) & 2 \\
Electron softness parameter, $\Lambda_e$
$(\mathrm{s}^{-1}\cdot\mathrm{J}^{-\Upsilon_e})$ & $3.6\times10^{18}$ \\
Hole softness parameter, $\Lambda_h$
$(\mathrm{s}^{-1}\cdot\mathrm{J}^{-\Upsilon_h})$ & $1.0\times10^{18}$ \\
Electron approaching index, $\Upsilon_e$ & 1 \\
Hole approaching index, $\Upsilon_h$ & 1 \\
\end{tabular}
\end{ruledtabular}
\end{table}

\subsection{Avalanche statistics}
For each bias point, $10^4$ injected carriers are used, consistent with the ensemble size used in prior AlInAsSb Monte Carlo work~\cite{karimi2025analysis}. The mean gain is obtained from the collected-carrier ensemble, and the excess-noise factor is
\begin{equation}
F=\frac{\langle M^2\rangle}{\langle M\rangle^2}.
\label{eq:noise}
\end{equation}

\section{Validation for A\lowercase{l}$_{0.7}$I\lowercase{n}A\lowercase{s}S\lowercase{b}}
The framework is tested against a \SI{1}{\micro\meter}-thick Al$_{0.7}$InAsSb p-i-n APD. The structure contains a \SI{30}{nm} GaSb cap, a \SI{100}{nm} p-type contact, a \SI{1}{\micro\meter} unintentionally doped multiplication layer, and a \SI{200}{nm} n-type contact on GaSb~\cite{dadey2023considerations,karimi2025analysis}. The layer dimensions and doping, together with the EDTB band parameters and the scattering-rate tables, define the device input. A uniform electric field is applied across the multiplication layer, following the approximation used in earlier AlInAsSb Monte Carlo simulations~\cite{zheng2020fullband,karimi2025analysis}.

\begin{figure}[t]
\includegraphics[width=0.85\columnwidth]{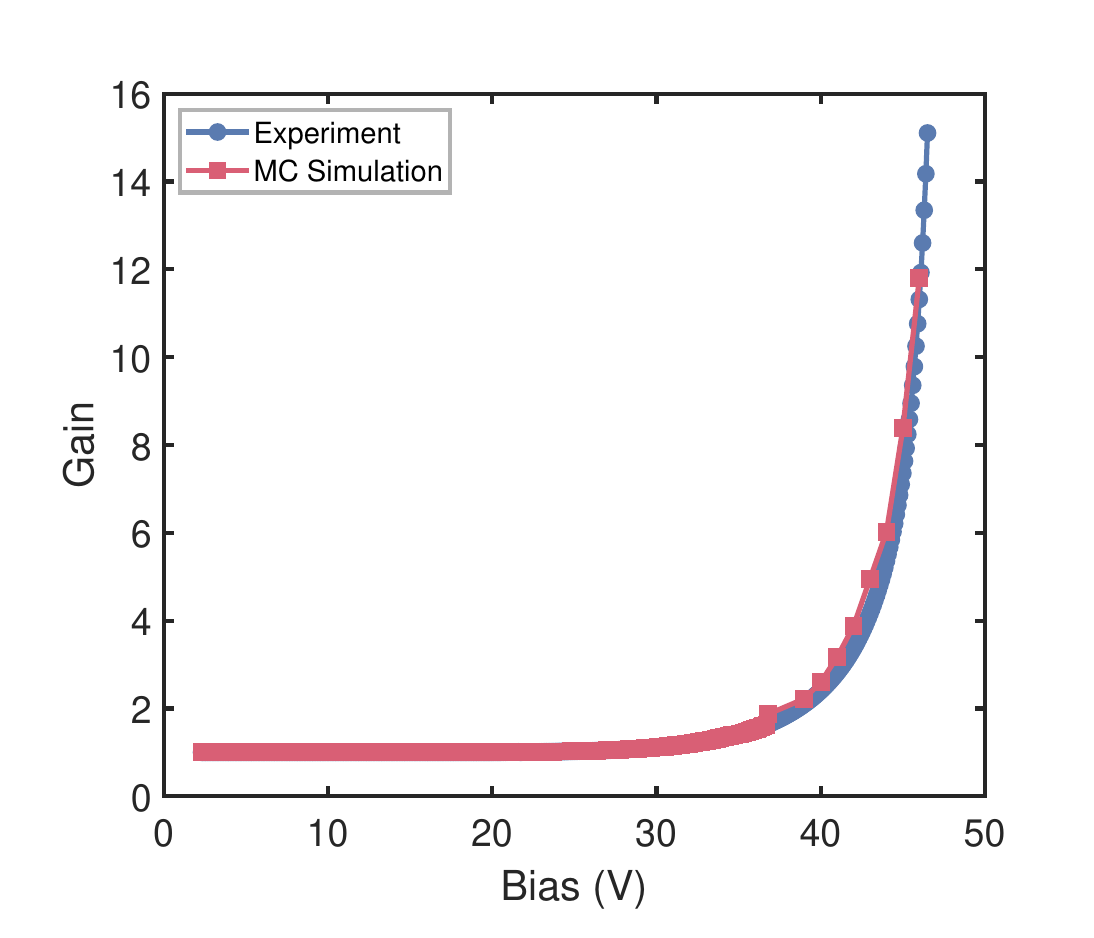}
\caption{Simulated (square symbols) and experimental (circular symbols) gain versus bias voltage for the \SI{1}{\micro\meter} Al$_{0.7}$InAsSb p-i-n APD.}
\label{fig:gain}
\end{figure}

\begin{figure}[htbp]
\includegraphics[width=0.85\columnwidth]{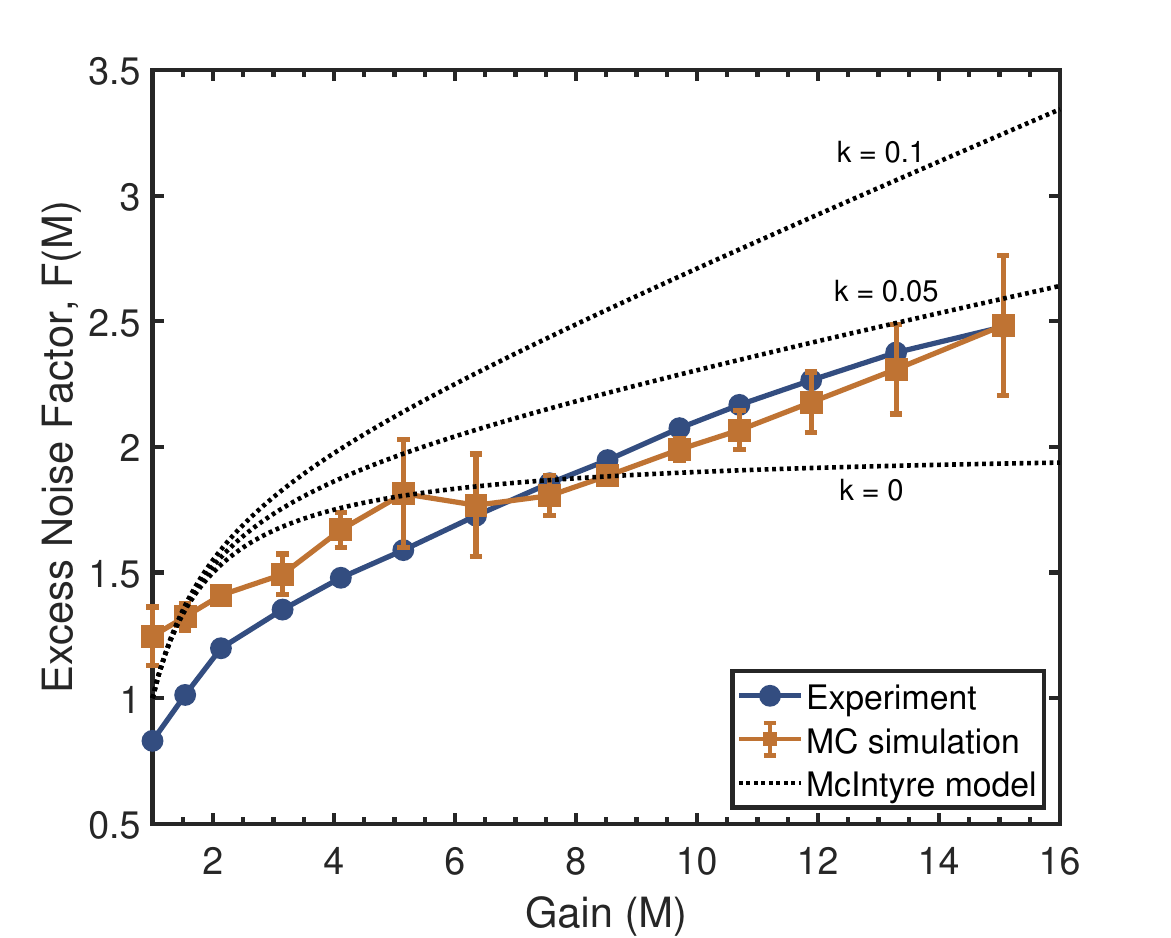}
\caption{Simulated (square symbols) and experimental excess-noise factor versus gain. The dashed curves show the McIntyre local-field model for the indicated ionization-coefficient ratios $k$.}
\label{fig:excessnoise}
\end{figure}

Figure~\ref{fig:gain} compares the calculated gain with the measured gain. The simulation reproduces the gradual increase at lower bias and the rapid rise near breakdown. Figure~\ref{fig:excessnoise} compares the excess-noise factor with measurement and with McIntyre curves. The measured and simulated values remain below the higher-$k$ local-field curves over the displayed gain range. Low excess noise has been reported across the AlInAsSb material system~\cite{woodson2016lownoise,jones2020lownoise,dadey2023considerations}.

Beyond gain and excess noise, the same Monte Carlo trajectories provide carrier-energy distributions, valley populations, and ionization-path statistics. These outputs can support mechanism studies without changing the transport engine. The present paper does not use them to assign a unique microscopic origin to the measured noise; the purpose here is to establish the modeling pipeline and its agreement with the device-level observables.

\subsection{Transferability and present limitations}
The workflow is not tied to one AlInAsSb composition. For another zinc-blende ternary or quaternary alloy, the same sequence can be followed: construct the atomistic structure, calculate the EDTB bands, build the scattering tables for the new composition, calibrate the impact-ionization model, and run the device Monte Carlo calculation. This is procedural transferability. It does not mean that one fitted Al$_{0.7}$InAsSb parameter set can be copied unchanged to another material.

Several limitations remain. First, the present device validation uses one Al$_{0.7}$InAsSb multiplication structure. Second, the electric field is treated as uniform in the UID layer rather than solved self-consistently with Poisson's equation. Third, the Keldysh impact-ionization law still contains calibrated parameters. Finally, the alloy-scattering calculation uses the random-alloy limit $S=1$ while the electronic structure is evaluated for a digital-alloy superlattice. These approximations are stated explicitly so that future extensions can test them independently. Comparison with measured ionization coefficients~\cite{yuan2019ionization}, additional multiplication-layer thicknesses, and other alloy systems would provide useful next tests of the framework.

\section{Conclusions}
We presented a full-band Monte Carlo workflow that connects an atomistic environment-dependent tight-binding model to composition-dependent scattering and avalanche transport in complex III--V alloys. For Al$_{0.7}$InAsSb, the EDTB calculation resolves the valley structure and digital-alloy band features used by the transport model. The scattering model includes alloy disorder, polar-optical, acoustic, intervalley-phonon, and ionized-impurity processes, while impact ionization is represented by a calibrated Keldysh rate. The calculated gain and excess noise of a \SI{1}{\micro\meter} Al$_{0.7}$InAsSb p-i-n APD agree with the measured trends. The main value of the framework is that the electronic structure and scattering inputs are generated through one documented sequence rather than assembled from a separate empirical parameter set for each new alloy. The same workflow can therefore be tested systematically for other ternary and quaternary APD materials.

\begin{acknowledgments}
This work was funded by National Science Foundation grant NSF 2430629.
\end{acknowledgments}

\bibliography{references}

\end{document}